# Challenges of Translating HPC codes to Workflows for Heterogeneous and Dynamic Environments

Fayssal Benkhaldoun, Christophe Cérin, Imad Kissami
LAGA-LIPN, University of Paris 13
99, avenue Jean-Baptiste Clément
93430 Villetaneuse, France
fayssal@math.univ-paris13.fr
Email:{christophe.cerin, imad}@lipn.univ-paris13.fr

Walid Saad
LaTICE: University of Tunis
ENSIT, 5 av. Taha Hussein, B.P. 56
Bab Mnara, Tunis, Tunisia
Email: walid.saad@lipn.univ-paris13.fr

**POSTER PAPER**

*Abstract*—In this paper we would like to share our experience for transforming a parallel code for a Computational Fluid Dynamics (CFD) problem into a parallel version for the RedisDG workflow engine. This system is able to capture heterogeneous and highly dynamic environments, thanks to opportunistic scheduling strategies. We show how to move to the field of "HPC as a Service" in order to use heterogeneous platforms. We mainly explain, through the CFD use case, how to transform the parallel code and we exhibit challenges to 'unfold' the task graph dynamically in order to improve the overall performance (in a broad sense) of the workflow engine. We discuss in particular of the impact on the workflow engine of such dynamic feature. This paper states that new models for High Performance Computing are possible, under the condition we revisit our mind in the direction of the potential of new paradigms such as cloud, edge computing…

## I. INTRODUCTION

In the field of High Performance Computing on clusters, researchers and programmers still continue to implement their applications through standards such as MPI or OpenMP. There also exists very good kernels in each scientific domains that are used as libraries in more general parallel frameworks. In this paper we address the following problem. Given a parallel code that uses kernels, how to transform it into a scientific workflow at a minimal cost? We do not want to explain and to teach to the programmers a new programming language but we rather prefer to teach them some kind of graphical language i.e. a task graph that represents the parallel code and, in the future the application to be developed. Then, automatically and after this translation, the task graph is executed by a workflow engine. This process opens new challenging questions from the workflow engine point of view, among them, how to combine different execution models for the tasks and what is the impact of such feature on the architecture of the workflow engine?

Our research focuses on the design of Systems for heterogeneous and highly dynamic environments, notably clouds, desktop grids and volunteer computing projects. The overall objective is to execute computational codes in such environments…and progressively moving from a traditional view for High Performance Computing (HPC) to Service oriented and workflow oriented views. A hard question, that the dynamicity causes here, is that given a workflow to schedule, we do not have any A-priori knowledge on the resources that are available. To address it, we proposed in [1] to implement a Publish-Subscribe [2] based mechanism for resource discovery and allocation. Recall that the Publish-Subscribe paradigm is an asynchronous mode for communicating between entities [2]. Some users, namely the subscribers or clients or consumers, express and record their interests under the form of subscriptions, and are notified later by another event produced by other users, namely the producers.

Indeed, we support the thesis that for building Systems for heterogeneous and highly dynamic environments we need to be compliant with:

  1. a publish-subscribe layer for the orchestration of the components of the system;

  2. a set of opportunistic strategies for allocating work/tasks that are also based on the publish-subscribe layer;

  3. a small number of software dependencies for the system and the ability to deploy the system and its applications on demand. This point is of particular interest in this paper and we promote the 'easy to use', and systems that can be deployed without a system administrator.

In this paper we examine the challenges and issues of coupling different execution models into the RedisDG workflow engine that has been developed according to the previous thesis. In other words, RedisDG is a concrete instance of the previous principles. We discuss of the impact on the thesis of such new features, and for that we consider a CFD problem. We execute our CFD solution, obtained after transforming a parallel code into a workflow, on top of the RedisDG system.

The organization of the paper is as follows. In section I we introduce the numerical problem we are faced to. We also summarize some related works. Section II introduces a parallel solution of the problem in the spirit of MPI programming.

Section III explains how to provide with a workflow oriented view for solving the numerical problem. Section IV explains the challenges we are faced to when we deal with multiple runtime systems for executing the nodes of the workflow. This analysis is the contribution of the paper. Section V concludes the paper.

## II. NUMERICAL PROBLEM

ADAPT [3] is an object oriented platform for running numerical simulations with a dynamic mesh adaptation strategy and coupling between finite elements and finite volumes methods. ADAPT, as a CFD (Computational Fluid Dynamics) software package, has been developed for realizing numerical simulation on an unstructured and adaptive mesh for large scale CFD applications. The ADAPT project is concerned with skills from numerical analysis, computer science and physics, and treats many physical phenomena such as combustion, plasma discharge, wave propagation, etc. The object oriented ADAPT platform is able to do coupling between finite volumes and finite elements methods. Yet another advantage of ADAPT is to focus in streamers which can be used in many applications (treatment of contaminated media, etching and deposition of thin films…). The key equation we solve is the evolution equation coupled with that Poisson equation:

$$\begin{cases} \frac{\partial U}{\partial t} + F(P,U) = S, \\ \Delta P = Q. \end{cases} \quad (1)$$

where F is discretized using the finite volume method on an unstructured mesh. The time-integration of the transport equation is performed using an explicit scheme. The discretized form of the Poisson equation leads to a linear system to be solved. Before our work, the existing ADAPT implementation was a sequential C++ code for each phenomenon, and the code requires a huge CPU time for executing a 3D simulation. For example, the 3D streamer code may run up to 30 days before returning results. The methodology we used for parallelization is based on the SPMD (single program, multiple data) paradigm. SPMD is the most common style of parallel programming and usually refers to message passing programming on distributed memory computer architectures.

## III. A PARALLEL SOLUTION

The pseudo code 1 shows how the parallelization is done for the streamer code according to the coupling of evolution equation with Poisson equation. We can see that most parts of the code are parallel ones (line 19, line 9 to 11 and line 22 to 26) except reading and splitting mesh in the beginning and between lines 14 and 15. Indeed, at this step, we construct the matrix that will be used to solve the linear system in line 19: the matrix is computed one time because it depends only on the mesh. We also put information about the overlapping between the communication and computation steps that are source for performance.

```
Algorithm 1: Algorithm of parallel ADAPT
 1  W:double;
 2  if rank==0 then  /* for the master
       processor                         */
 3  |   Read mesh data;
 4  |   Split mesh with METIS;
 5  |   Distribute mesh to all processors;
 6  end
 7  W=0;
 8  for each rank do  /* for each processor */
 9  |   Initialize conditions and create constants;
10  |   Send the information of halos cells to neighbor
        sub-domains;
11  |   Apply boundary conditions;
12  end
13  if rank==0 then
14  |   Construct matrix of linear system;
15  |   Split the matrix and send part of each processor;
16  end
17  for each iteration do
18  |   for all rank do  /* for all processors
           */
19  |   |   Solve linear system using MUMPS;
20  |   end
21  |   for each rank do
22  |   |   Send the information of halos cells to
            neighbor sub-domains;
23  |   |   Apply boundary conditions;
24  |   |   Compute fluxes of convection, diffusion and
            source term;
25  |   |   Update solution : $W^{n+1} = W^n + \Delta t *$
            $(rez\_conv + rez\_dissip + rez\_source)$;
26  |   |   Save results in parallel way using Paraview;
27  |   end
28  end
```

## IV. A WORKFLOW SOLUTION

### A. A brief introduction to scientifique workflows

Applications in e-Science are becoming increasingly large-scale and complex. These applications are often in the form of workflows [4] such as MONTAGE, Blast [5], CyberShake [6] with a large number of software components and modules. Workflow and scheduling policies have been studied for one decade or two. In the reminder of this section we consider many categories of strategies. At a high level of abstraction, we focus first on the works that seek to optimize the execution time and/or QoS constraints of the workflows running in grid environments, and second works anchored in the Map-Reduce framework (data aware scheduling). Workflow management systems related works are presented in the synthesis from Valduriez [7] or in the work of Carole Goble [8]. These two papers are more related to cloud computing and data intensive scientific workflows in putting an emphasis on data management. The context of these works is not adapted to our context because they do not take into account a dynamic view of the system in reacting to events when they arrive. Static information are supposed to be available (task graph, date of the events, task duration and costs…).prepared text file. You are now ready to style your paper; use the scroll down window on the left of the MS Word Formatting toolbar.

## B. Transformation of the parallel code into a workflow

From a methodological point of view, we have realized a conversion of the parallel ADAPT algorithm described with algorithm 1, page 2 into a Direct Acyclic Graph (DAG). The ADAPT graph that we schedule has the shape depicted on

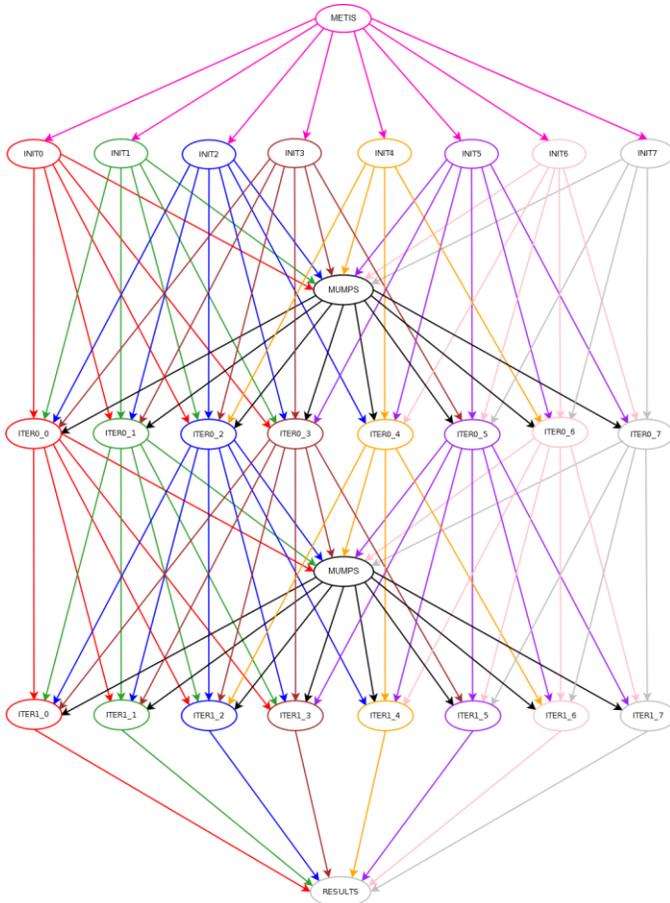

Fig. 1. The ADAPT workflow

Figure 1. As a first approximate, each node of the DAG corresponds to a line in the algorithm 1, page 2. For instance, the node labeled METIS corresponds to line 4, the line 9 corresponds to the nodes INIT{0..7}, the node label-led MUMPS corresponds to line 19, the lines 21 to 27 correspond to the nodes ITER{0..number of iteration}_{0..7}. The DAG describes parallel execution, for instance nodes INIT{0..7} may run in parallel if we have enough processors. The DAG also describes dependencies e.g. precedence relations, between nodes: nodes INIT{0..7} may be executed only after the completion of METIS.

The key idea is to decompose the solution according to services (METIS, MUMPS…) and let them being scheduled by the RedisDG workflow engine. The key idea is no more with the design of tightly coupled parallel codes, as with MPI, but rather, to adopt a higher view in terms of general services. As with the design of computer programs, the process of converting a parallel or sequential program into a scientific workflow is not a formal process and it is deeply anchored into the experience of the programmer in term of idioms and best practices.

## V. CHALLENGE IN USING MULTIPLE EXECUTION MODELS

### A. The RedisDG workflow engine

We first introduce our workflow engine, its key components. Second, we analyze the problems of coupling multiple execution models, motivated by the will to deploy them on demand, because our target environments are heterogeneous and highly dynamic. The key idea is to adopt an opportunistic point of view for executing tasks: when a request for executing a task happens, we examine the situation and the knowledge at our disposal, and then we take a decision.

A scientific workflow system is a specialized form of a workflow management system designed specifically to compose and execute a series of computational or data manipulation steps, or workflow, in a scientific application. In this thesis 'workflow' and 'scientific workflow' are considered equivalent for sake of simplicity. The simplest computerized scientific workflows are scripts that call in data, programs, and other inputs and produce outputs that might include visualizations and analytic results. These may be implemented in programs such as R or MATLAB, or using a scripting language such as Python or Perl with a command-line interface. By focusing on the scientists, the focus of designing scientific workflow system shifts away from the workflow scheduling activities, typically considered by grid computing environments in the past and now by cloud computing environments for optimizing the execution of complex computations on predefined resources, to a domain-specific view of what data types, tools and distributed resources should be made available to the scientists and how can one make them easily accessible and with specific Quality of Service (QoS) requirements.

RedisDG protocol: in Figure 2, we present the RedisDG architecture and we now introduce the steps of an application execution. In RedisDG, a task may have five states: WaitingTasks, TasksToDo, TasksInProgress, TasksToCheck and FinishedTasks. These states are managed by five actors: a broker, a coordinator, a worker, a monitor and a checker. Taken separately, the behavior of each component in the system may appear simple, but we are rather interested in the coordination of these components, which makes the problem more difficult to solve. The key idea is to allow the connection of dedicated components (coordinator, checker,…) in a general coordination mechanism in order to avoid building a monolithic system. The behavior of our system as shown in Figure 2 is as follows:

1. Tasks batches submission. Each batch is a series-parallel graph of tasks to execute.

2. The Broker retrieves tasks and publishes them on the channel called *WaitingTasks*.

3. The Coordinator is listening on the channel *WaitingTasks*.

4. The Coordinator begins publishing independent tasks on the channel *TasksToDo*.

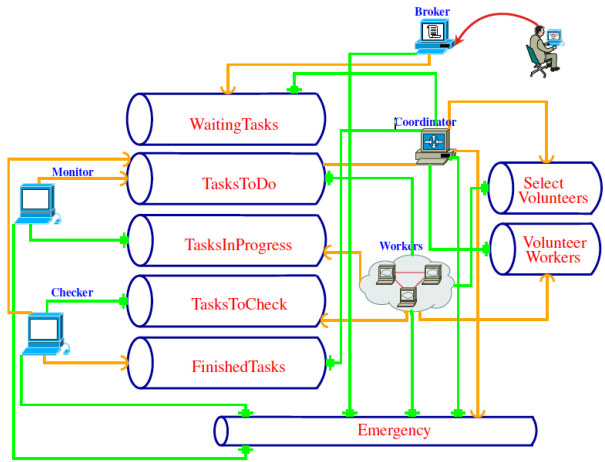

Fig. 2. Interactions between components of the RedisDG system

5. Workers announce their volunteering on the channel *VolunteerWorkers*.

6. The coordinator selects Workers according to SLA criteria.

7. The Workers, listening beforehand on the channel *TasksToDo* start executing the published tasks. The event 'execution in progress' is published on the channel *TasksInProgress*.

8. During the execution, each task is under the supervision of the Monitor whose role is to ensure the correct execution by checking if the node is alive. Otherwise the Monitor publishes again, tasks that do not arrive at the end of their execution. It publishes, on the channel *TasksToDo*, in order to make the execution of the task done by other Workers.

9. Once the execution is completed, the Worker publishes the task on channel *TasksToCheck*.

10. The Checker verifies the result returned and publishes the corresponding task on the channel *FinishedTasks*.

11. The Coordinator checks dependencies between completed tasks and those waiting, and restarts the process in step (4).

12. Once the application is completed (no more tasks), the Coordinator publishes a message on the channel *Emergency* to notify all the components by the end of the process.

### B. Motivations for using multiple execution models

As big experiments manage large amounts of computation and data, it becomes critical to execute them in high-performance computing environments, such as clusters, grids, and clouds. However, few workflow systems provide parallel support and they usually need labor-intensive work, with limited gain, through primitives to optimize workflow execution. The needs are to specify and to enable the optimization of parallel execution of scientific workflows.

One important issue is to describe a collaborative system able to execute the tasks graph (DAG), more precisely a hierarchical DAG because a node needs to be unfolded into different forms of execution runtime. In this vein, we can imagine to unfold a node, dynamically, if we have enough resource, or we could request, on demand, new resources. At an abstract level, such systems are usually modeled with centralized and state-based formalisms like automata, Petri nets or state-charts. They can also directly be specified with dedicated notations like BPEL [9] or BPMN [10]. In this context we will surely also need the contributions from the fields of lazy evaluation, abstract grammar and context-free languages as well as from software engineering in a broad sense. We now analyze some relevant works in this context.

#### 1) OpenAlea

The OpenAlea system [11], [12] is a workflow system based on $\lambda$-dataflow which is able to uniformly deal with classical data analysis, visualization, modeling and simulation tasks. $\lambda$-dataflow means that the system uses higher-order constructs in the context of dataflows theory and thus allows to represent control flow using algebraic operators (e.g., conditionals, map/reduce…). An actor in OpenAlea is an elementary brick (a.k.a. component or activity) that has a name, a function object (a functor, a program, a web service, a micro service or a composite actor), and explicitly defined input and output ports. A semantic type is associated to each port (with a corresponding color). A workflow is represented as a directed multi-graph where nodes are actors, and directed edges are data links between output and input ports. A workflow can become a (composite) actor in another workflow to allow composition. In another word, one of the main originality of OpenAleat is to introduce higher-order dataflows as a means to uniformly combine classical data analysis with modeling and simulation.

Another major originality of OpenAlea lies in the way iteration is handled by introducing a specific kind of actor, called dataflow variable X. It allows to specify that, at a given port, an actor receives an unbound variable rather than a value. Connecting an X to an actor transforms a workflow into a lambda function, and allows to express higher-order programming providing control flow behavior using a set of algebraic operators. The three iteration types can be expressed as (1) counting loops without dependencies (map operator), (2) counting loops with dependencies (reduce and for operators) and (3) conditional loops (while operator). As we can observe, the orientation of the framework is clearly towards data-intensive jobs since we recognize some operations of the popular MapReduce framework.

#### 2) Chiron

Chiron [13] is a workflow execution engine designed to run workflows in parallel in HPC environments. Chiron uses the message passage interface (MPI) so that the engine is executed along the computing nodes of the environment. Each computing node runs an instance of Chiron, which also gathers provenance data (start time, end time, execution status, logs, errors). This last point is a strong feature. Provenance is a key element to assess the correctness of the experiment and its reproducibility. Through provenance, scientists can follow the experiment execution and verify, for example, which parameter values produced the best results.

Chiron addresses the issue of optimizing parallel workflow execution, according to a specific algebra for scientific workflows. This algebra is inspired by the relational algebra for databases and provides a uniform data model that expresses all experiment data as relations. For Chiron, a scientific workflow is then a set of algebraic expressions. Basically, an operator applies on an activity which is a program or an expression (plus an input relation schema as well as an output relation schema), and produces a relation in the sense of relational database.

The proposed scientific workflow algebra of Chiron considers six different operators. There are for instance four algebraic operators that invoke computer programs. They differ basically in the way tuples are consumed and produced. For instance, the Map operator rules activities that, for each input tuple, produce a single output tuple. This is typically the most general case because most computer programs consume a set of input parameter values to produce a set of output parameter values. Another example is the SplitMap operator which is related to fragmentation and decomposition methods which, based on a single tuple, may produce several output tuples.

Each node has an instance of Chiron, and each instance has a thread to schedule activations to the available activation "processors threads" in the instance. In node 0, there is a thread called "workflow processor", which orchestrates the workflow execution deciding what is ready to be consumed. The "activation schedulers" use MPI to communicate with the "workflow processor" thread. This communication is used for reporting completed activations and obtaining new activations to be consumed. Whenever a given activation runs in blocking mode, the workflow processor answers a request with a "wait" message instead of answering with an "activation ready" to be consumed.

*3) A component model for HPC applications*

In [14] authors study the feasibility of efficiently combining both a software component model and a task-based model. Task based models are known to enable efficient executions on recent HPC computing nodes while component models ease the separation of concerns of application and thus improve their modularity and adaptability. This paper is a Software Engineering effort for capturing maintainability of HPC codes. Authors notice that HPC task-based scheduling runtime systems have been designed to ease reaching high performance on complex hardware as well as performance portability. They keep the option that task granularity should be small enough such that the runtime scheduling algorithm can leverage the flexibility to make the efficient choices for the available hardware. In others words they assume that overlapping or combining small computational chunks is better for performance.

Then authors introduce the COMET programming model is based on the $L^2C$ model, a minimal HPC-oriented component model [15]. The COMET runtime distinguishes three types of components: a) components written by the user in the programming model, b) components generated during the compilation phase, and c) components written by experts that make the runtime easily extensible to support potential new concepts in the programming model. The task implementation is provided by a use of a port of the "meta-task" that is connected to a component instance outside any dataflow section. Hence, meta-tasks do not contain user-level code; they only enable the implicit description of task sets; the actual implementation is delegated to components. The expression of task parallelism is achieved thanks to the composition of meta-tasks of dataflow sections, while data parallelism is expressed inside meta-task using data partitioning and alignment expressions.

*C. Challenges and possible adaptation of RedisDG*

Summarizing the previous subsection, we can say that three directions have been explored, all of them put an emphasis on the notion of a task graph, but for objectives that are quite different but important nowadays: performance (OpenAlea), reproducibility and provenance (Chiron), maintainability (COMET). We propose to keep some features for the RedisDG workflow engine and we assume that only the workflow description (currently implemented as an XML file) can be enriched. We do not want to count on "external" tools or techniques; we mean for instance the introduction of a new programming model. The programming model is the DAG, that's it!

We now introduce an example to explain how we plan to go through "data life cycle" (DLC) and interoperability with other execution models (EM). Data life cycle is the description of operational stages through which data pass when we enter to a system and until we leave the system. We have in mind the Active Data[1] and the StarPU[2] frameworks and we attempt to offer some kind of unified view. The challenges are to expose an high level view for DLC and EM across distributed systems and architectures and also to expose interactions between the infrastructure and DLC/EM. For instance we need to react when a failure occurs in data transmission or we need to configure an execution model when we detect its presence. Regarding the data transmission problem, the system should drop the whole dataset (for instance the input files of a file), remove any associated file and metadata, re-acquire the dataset using the same parameters. Regarding the execution model, the system should interact with the physical node to detect if the environment variable for the number of logical GPU devices is lower than the number of physical devices. The system should also check the default scheduling policy and the availability of a performance model and changes them according to the data that will be computed on the node.

For the modeling, we propose a collaborative system which is centered on the notion of user's workspace. We assume that the workspace of a user is given by a *map*. It is a tree used to visualize and organize tasks in which the user is involved together with information used for the resolution of the tasks. The workspace of a given user may, in fact, consist of several maps where each map is associated with a particular service offered by the user. To simplify, one can assume that a user offers a unique service so that any workspace can be identified with its graphical representation as a map. Each map is associated with a node of the DAG and it specifies the resolution of the task regarding DLC/EM. The following map might represent the workspace of a task regarding our toy example:

```
Task_i --- DLC --- ?transmission --- drop -- remove -- reload
       |
       |- EM  --- ?GPU -- fix_number_devices
              |
              |- ?Scheduling --- data_aware --- choose_data_aware
                              |
                              |- in_memory  --- choose_in_memory
```

---

1. http://graal.ens-lyon.fr/~gfedak/pmwiki-test/pmwiki.php/Main/ActiveData
2. http://starpu.gforge.inria.fr/doc/html/index.html

We interpret a task as a problem to be solved, that can be completed by refining it into sub-tasks using some kind of *business rules*. In a first approximation a business rule can be modeled as a production rule $P : s_0 \rightarrow s_1 \rightarrow \cdots \rightarrow s_n$ stating that task $s_0$ can be reduced to subtasks $s_1$ to $s_n$. We argue that the Guarded Attribute Grammar Syntax, as in , is a good formal model to continue into this direction.

However, to become concrete, the modeling implies that we revisit also the RedisDG protocol. At least two new communication channels (one for capturing DLC interactions, and one for capturing EM interactions) needs to be added. These channels connect the coordinator, the worker actors but could also impact the broker since in some cases we may need to re-publish a task (see Figure 2). Future work needs to be done to decide on the modifications.

VI. CONCLUSION

In this paper we introduced our methodology, through an example, for designing and solving computational problem inside the ADAPT framework according to a Service oriented view, i.e. according to requirements we found for clouds, with the objective to exploit heterogeneous platforms and highly dynamic environments. The basic idea is to admit that excellent, in terms of performance, numerical libraries are now available because they have been optimized for many hardware architectures. The problem is not so much to try again to improve performance but rather, in our opinion, to exploit all the resources in terms of computing, networking, storing, available on the planet and/or in major cloud providers and for everyone.

One opportunity we analyzed in this paper is the following. In our past work, we have assumed, implicitly, that each node corresponds to a sequential task. The parallelism is depicted by parallel edges of the workflow and started from the same node. We could imagine that a node corresponds to a parallel execution, for instance a parallel version of METIS or MUMPS for instance. The related work for capturing such feature exhibits many possibilities but, none of them are good enough to capture heterogeneous and highly dynamic environments as with our RedisDG system. We isolated the key challenges to address in order to adapt the RedisDG system for capturing Data Life Cycle (DLC) and Execution Model (EM) requirements.

REFERENCES

1. Leila Abidi, Jean-Christophe Dubacq, Christophe Cérin, and Mohamed Jemni. A publication-subscription interaction schema for desktop grid computing. In Sung Y. Shin and José Carlos Maldonado, editors, SAC, pages 771–778. ACM, 2013.
2. Patrick Th. Eugster, Pascal Felber, Rachid Guerraoui, and Anne-Marie Kermarrec. The many faces of publish/subscribe. ACM Comput. Surv.,35(2):114–131, 2003.
3. Fayssal Benkhaldoun, Jaroslav Fort, Khaled Hassouni, and Jan Karel. Simulation of planar ionization wave front propagation on an unstructured adaptive grid. Journal of Computational and Applied Mathematics, 236(18):4623 – 4634, 2012. {FEMTEC} 2011: 3rd International Conference on Computational Methods in Engineering and Science, May 9–13, 2011. URL: http://www.sciencedirect.com/science/article/pii/S0377042712001707, doi:http://dx.doi.org/https://doi.org/10.1016/j.cam.2012.04.010.
4. S. Bharathi, A. Chervenak, E. Deelman, G. Mehta, Mei-Hui Su, and K. Vahi. Characterization of scientific workflows. In Workflows in Support of Large-Scale Science, 2008. WORKS 2008. Third Workshop on, pages 1–10, Nov 2008. doi:10.1109/WORKS.2008.4723958.
5. Stephen F. Altschul, Thomas L. Madden, Alejandro A. Schäffer, Jinghui Zhang, Zheng Zhang, Webb Miller, and David J. Lipman. Gapped blast and psiblast: a new generation of protein database search programs. NUCLEIC ACIDS RESEARCH, 25(17):3389–3402, 1997.
6. Robert Graves, Thomas H. Jordan, Scott Callaghan, Ewa Deelman, Edward Field, Gideon Juve, Carl Kesselman, Philip Maechling, Gaurang Mehta, Kevin Milner, David Ok aya, Patrick Small, and Karan Vahi. CyberShake: A Physics-Based Seismic Hazard Model for Southern California. Pure and Applied Geophysics, 168:367–381, 2011. doi:10.1007/s00024-010-0161-6.
7. Ji Liu, Esther Pacitti, Patrick Valduriez, and Marta Mattoso. A survey of data-intensive scientific workflow management. J. Grid Comput., 13(4):457–493, 2015. URL: http://dx.doi.org/10.1007/s10723-015-9329-8, doi:10.1007/s10723-015-9329-8.
8. Yogesh Simmhan, Lavanya Ramakrishnan, Gabriel Antoniu, and Carole A. Goble. Cloud computing for data-driven science and engineering. Concurrency and Computation: Practice and Experience, 28(4):947–949, 2016. URL: http://dx.doi.org/10.1002/cpe.3668, doi:10.1002/cpe.3668.
9. Arnaud Lanoix, Julien Dormoy, and Olga Kouchnarenko. Combining proof and model-checking to validate reconfigurable architectures. Electronic Notes in Theoretical Computer Science, 279(2):43–57, 2011.
10. Jeffrey O Kephart and David M Chess. The vision of autonomic computing. Computer, 36(1):41–50, 2003.
11. Christophe Pradal, Samuel Dufour-Kowalski, Frédéric Boudon, Christian Fournier, and Christophe Godin. Openalea: a visual programming and component-based software platform for plant modelling. Functional Plant Biology, 35(10):751–760, 2008. URL: http://dx.doi.org/10.1071/FP08084.
12. Christophe Pradal, Christian Fournier, Patrick Valduriez, and Sarah Cohen Boulakia. Openalea: scientific workflows combining data analysis and simulation. In Proceedings of the 27th International Conference on Scientific and Statistical Database Management, SSDBM '15, La Jolla, CA, USA, June 29 - July 1, 2015, pages 11:1–11:6, 2015. URL:http://doi.acm.org/10.1145/2791347.2791365; doi:10.1145/2791347.2791365.
13. Eduardo S. Ogasawara, Jonas Dias, Vítor Silva Sousa, Fernando Seabra Chirigati, Daniel de Oliveira, Fábio Porto, Patrick Valduriez, and Marta Mattoso. Chiron: a parallel engine for algebraic scientific workflows. Concurrency and Computation: Practice and Experience, 25(16):2327–2341, 2013. URL: http://dx.doi.org/10.1002/cpe.3032, doi:10.1002/cpe.3032.
14. Olivier Aumage, Julien Bigoty, Hélène Coullon, Christian Pérez, and Jérôme Richard. Combining both a component model and a task-based model for hpc applications: a feasibility study on gysela. In 17TH IEEE/ACM INTERNATIONAL SYMPOSIUM ON CLUSTER, CLOUD AND GRID, CCGRID '17, pages ??–??, IEEE, Piscataway USA, 2017. IEEE. Julien Bigot, Zhengxiong Hou, Christian Pérez, and Vincent Pichon. A low level component model easing performance portability of HPC applications. Computing, 96(12):1115–1130, 2014. URL:http://dx.doi.org/10.1007/s00607-013-0368-3, doi:10.1007/s00607-013-0368-3.
15. Eric Badouel, Loïc Hélouët, Georges-Edouard Kouamou, Christophe Morvan, and Nsaibirni Robert Fondze, Jr. Active workspaces: Distributed collaborative systems based on guarded attribute grammars. SIGAPP Appl. Comput. Rev., 15(3):6–34, October 2015. URL: http://doi.acm.org/10.1145/2835260.2835261, doi:10.1145/2835260.2835261